\documentclass[twocolumn,twoside,slac_two]{revtex4}
\usepackage{graphicx}
\usepackage{fancyhdr}
\def\ifmath#1{\relax\ifmmode#1\else$#1$\fi}
\def\CP     {\ifmath{C\!P}}
\def\babar{\mbox{\sl B\hspace{-0.4em} {\scriptsize\sl A}\hspace{-0.4em} B\hspace{-0.4em} {\scriptsize\sl A\hspace{-0.1em}R}}}

\def\BB    {\ifmath{B\Bbar}} %<<<
\def\Kbar{\ifmath{\kern 0.2em\overline{\kern -0.2em K}}{}}   %<<<new
\def\Bbar{\ifmath{\kern 0.18em\overline{\kern -0.18em B}}{}} %<<<new
\def\Dbar{\ifmath{\kern 0.2em\overline{\kern -0.2em D}}{}}   %<<<new
\def\Y#1S{\ifmath{\Upsilon\rm(#1S)}} %<<<new

\def\UT{Unitarity Triangle}

\def\kev  {\ifmath{\mbox{\,ke\kern -0.08em V}}} %<<<
\def\mev  {\ifmath{\mbox{\,Me\kern -0.08em V}}} %<<<
\def\gev  {\ifmath{\mbox{\,Ge\kern -0.08em V}}} %<<<
\def\gevc {\ifmath{\mbox{\,Ge\kern -0.08em V$\!/c$}}} %<<<
\def\mevc {\ifmath{\mbox{\,Me\kern -0.08em V$\!/c$}}} %<<<
\def\gevcc{\ifmath{\mbox{\,Ge\kern -0.08em V$\!/c^2$}}} %<<<
\def\mevcc{\ifmath{\mbox{\,Me\kern -0.08em V$\!/c^2$}}} %<<<

\def\fb   {\ifmath{\mbox{\,fb}^{-1}}}

\newcommand{\DSS} {\ensuremath{D^{**}}}

\newcommand{\DI}  {\ensuremath{D_1}}
\newcommand{\DII} {\ensuremath{D_2^*}}

\newcommand{\DS}  {\ensuremath{D^{*}}}

\newcommand{\DZ}  {\ensuremath{D^0}}

\newcommand{\D}   {\ensuremath{D}}

\newcommand{\PIZ} {\ensuremath{\pi^0}}
\newcommand{\PIC} {\ensuremath{\pi^\pm}}

\newcommand{\PI}  {\ensuremath{\pi}}

\newcommand{\TO}  {\ensuremath{\to}}

\newcommand{\BR}     {\ensuremath{\mathcal B}}

\newcommand{\VCB}{\ensuremath{|V_{cb}|}}
\newcommand{\VUB}{\ensuremath{|V_{ub}|}}
\begin{document}

\title{Experimental Results on \VCB\ and $b\to c\ell\nu$ Transitions}


\author{A. Hauke}
\affiliation{Fachbereich E5a, Universit\"at Dortmund, Otto-Hahn-Str. 4, 44221 Dortmund, Germany}

\begin{abstract}
A review of recent analyses on semileptonic decays of $B$ mesons into charmed final states
is given. \VCB\ is extracted both by the \babar\ and the Belle collaboration from their
datasets using inclusively and exclusively reconstructed final states.

In addition there are recent results on the determination of exclusive branching fractions
to the ground states $D$ and $D^*$, as well as to excited $D^{**}$ states. Those play an
important role in understanding the composition of the total decay width. They represent
also a sizable fraction of the backgrounds for exclusive analyses and are presented here
as well.
\end{abstract}

\maketitle

\thispagestyle{fancy}


\section{Introduction}

Beside being one of the input parameters to the standard model, \VCB\ is one of the keys to
understand flavor physics and \CP-violation for two reasons. First, $|V_{ub}/V_{cb}|$ is
one of the sides of the \UT\ and second, the dominant $b \to c \ell\nu$ transitions give
a large background to any analysis aiming to measure \VUB. Hence understanding those decays
and the composition of the total decay width $\Gamma_{b\to c\ell\nu}$ is crucial to answer
the question if the CKM-mechanism of the standard model is the only and correct way to
describe \CP-violation.

To extract \VCB\ semileptonic decays are the best tool. On tree level the quark transition
$b \to c \ell\nu$ factorizes into an hadronic and a leptonic current and the vertex is
proportional to \VCB. Theoretical calculations of those currents are uncomplicated since there
are no corrections from strong final state interactions. From an experimentalists point of view
the high energetic lepton can be identified easily giving a good handle to separate the decays
of interest from backgrounds, even if the accompanying neutrino limits the knowledge of the
kinematics of the final state.

However we are unable to study the $b \to c$ transitions on quark level and the effects of the
strong interaction inside the hadrons is not yet completely understood theoretically. So
determining \VCB\ is always an interplay of theory calculations in the framework of QCD and
appropriate measurements whose results can be interpreted within these calculations.
In general there are two ways to do this:
\begin{itemize}
\item Exclusive measurements of a single hadronic final state, e.g. the ground state
\D\ or \DS, restrict the dynamics of the process. The remaining degrees of freedom,
usually connected to different helicity states of the charmed hadron, can be expressed in
terms of some (few) form factors, depending on the momentum transfer of the process.
The shapes of those form factors are unknown but can be measured. However, the overall normalization
of these functions need to be determined from theoretical calculations.
\item The opposite approach is to do an inclusive measurement of all possible final states.
This way all theory parameters can be adjusted to the measurement in a combined fit including
\VCB\ as one of the fit parameters. Usually the adjustment of theory and measurement is done
using the moments of some kinematic variables, e.g. the lepton energy $E_e$ or the hadronic mass.
\end{itemize}
For both approaches there are recent measurements from the two $B$ factories, Belle and \babar.

\section{Exclusive Semileptonic Decays}

Picking an exclusive decay mode in order to measure \VCB, decays to \DS\ are the ones to choose.
They have the largest branching fraction, experimentally a clean signature due to the sharp
resonance of the \DS\ in the mass difference $m(D\pi)-m(D)$, and for the theoretical description,
there are no corrections of order $1/m$ from heavy quark symmetry breaking due to Luke's Theorem.

\subsection{\VCB\ from Decays $B\to D^*\ell\nu$}

\begin{figure}
\includegraphics[width=80mm]{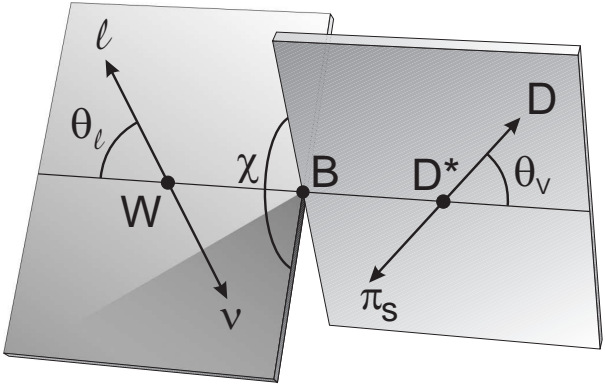}
\caption{\label{fig:DSexcl:angles}
Definition of kinematic quantities describing a decay $B\to \DS\ell\nu$.}
\end{figure}

The full kinematic of a decay $B\to\DS\ell\nu$ is described by three angular quantities: The two
decay angles $\theta_V$ and $\theta_\ell$ of the \DS\ and the virtual $W$ and the angle $\chi$
between those two decay planes. This situation is illustrated in Figure \ref{fig:DSexcl:angles}.
Together with the momentum transfer $q^2$ thus the full decay rate is
expressed differentially as
\begin{displaymath}
\frac{d\Gamma}{dq^2\,d\cos\theta_\ell\,d\cos\theta_V\,d\chi}
\end{displaymath}
which is given by the sum of three helicity amplitudes, $H_i$, representing the three possible
polarizations of the \DS.

Those $H_i$ are theoretically described by three form factors $R_1$, $R_2$, and $h_{A1}$ which are
functions of $w= v_B\cdot v_{\DS}$, the product of the four-velocities of the $B$ and the \DS.
Using the calculation by Caprini, Lellouch and Neubert \cite{DSexcl:caprini-ff} these functions are
expressed as
\begin{eqnarray}
R_1(w) &=& R_1(1) - 0.12(w-1) + 0.05 (w-1)^2 \nonumber \\
R_2(w) &=& R_2(1) + 0.11(w-1) - 0.06 (w-1)^2 \nonumber \\
h_{A1}(w) &=& h_{A1}(1) [ 1-\rho^2 z + (53 \rho^2 -15 ) z^2 \nonumber \\
          &&  - ( 231 \rho^2 -91 ) z^3 ] \nonumber
\end{eqnarray}
where $z$ is given as $z=\frac{\sqrt{w+1}-\sqrt{2}}{\sqrt{w+1}+\sqrt{2}}$ and $\rho$ is a
slope parameter, defining the shape of $h_{A1}$ away from the limit $w=1$.

To measure the differential decay rate, \babar\ selects decays $B^0 \to D^{*-}\ell\nu$ based on a
sample of 79\fb. Electrons or muons are required to have momenta of more than 1.2\gevc\ and the
\DS\ is reconstructed in the decay $D^{*-}\to \DZ\pi$ with $\DZ\to K\pi, K\pi\PIZ, K\pi\pi\pi$.

The most powerful variable to discriminate between signal and background is
\begin{displaymath}
\cos\theta_{B,D^*\ell} = - \frac{2 E_B E_{D^*\ell} - m_B^2 -m_{D^*\ell}^2}{2 |\vec p_B| |\vec p_{D^*\ell}|}.
\end{displaymath}
This gives the cosine of the angle between the momenta of the $B$ and the $\DS\ell$-system if, and only if
the $\DS\ell$ contains all decay products of the $B$ except a single massless particle. For signal events
this condition is fulfilled and the values of $\cos\theta_{B,D^*\ell}$ take on physical values, while for
background events the underlying conservation of four-momentum is not given, leading to calculated absolute 
values greater than one. Allowing for resolution effects events are selected with the requirement
$|\cos\theta_{B,D^*\ell}|<1.2$.

\begin{figure}
\includegraphics[width=80mm]{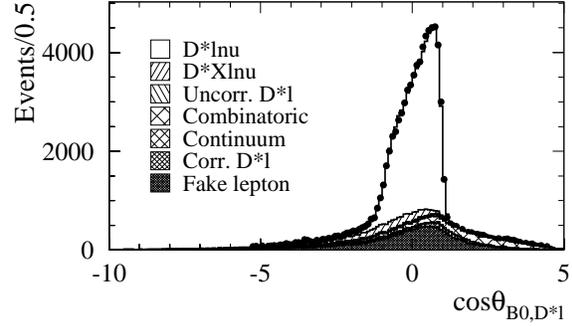}
\caption{\label{fig:DSexcl:cosby}
Distribution of the variable $\cos\theta_{B,D^*\ell}$ for data (dots) and simulation (histograms).}
\end{figure}

Figure \ref{fig:DSexcl:cosby} shows the distribution of the variable $\cos\theta_{B,D^*\ell}$ for data
and Monte Carlo. Clearly visible is the restriction for signal events to the allowed region $[-1,+1]$,
while for background events the distributions are smeared out to the unphysical region.

Since the direction of the $B$ in the center-of-mass frame is only known in magnitude but not in direction,
$w$ can not be reconstructed from the decay. Instead an estimator is calculated as the mean of the minimally
and maximally allowed values of $w$ that are still in agreement with four-momentum conservation of the decay.
This procedure gives a resolution for $w$ of about 0.4.

In principle one now should bin the data four dimensionally in $w$, $\theta_V$, $\theta_\ell$ and $\chi$,
but with the given size of the dataset this is not possible. Instead the data is projected on the three variables
$w$, $\cos \theta_V$ and $\cos \theta_\ell$. $\chi$ turns out to have the least significance on \VCB\ and is
neglected here, while the projections on the other variables are split into 10 bins each. The correlations
between them are taken into account when determining \VCB\ and the three form factors in a combined fit.

\begin{figure*}
\includegraphics[width=80mm]{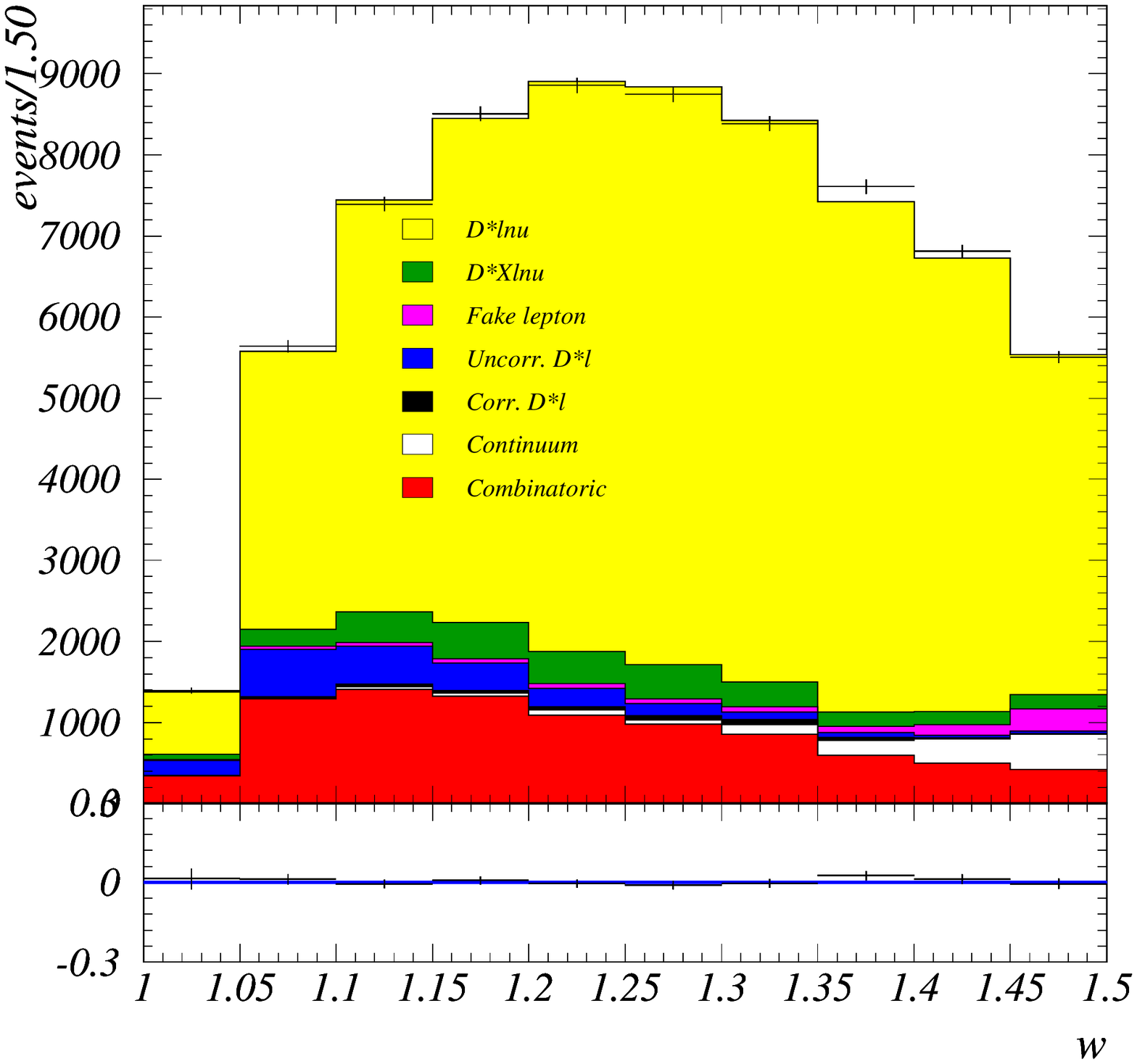}\includegraphics[width=80mm]{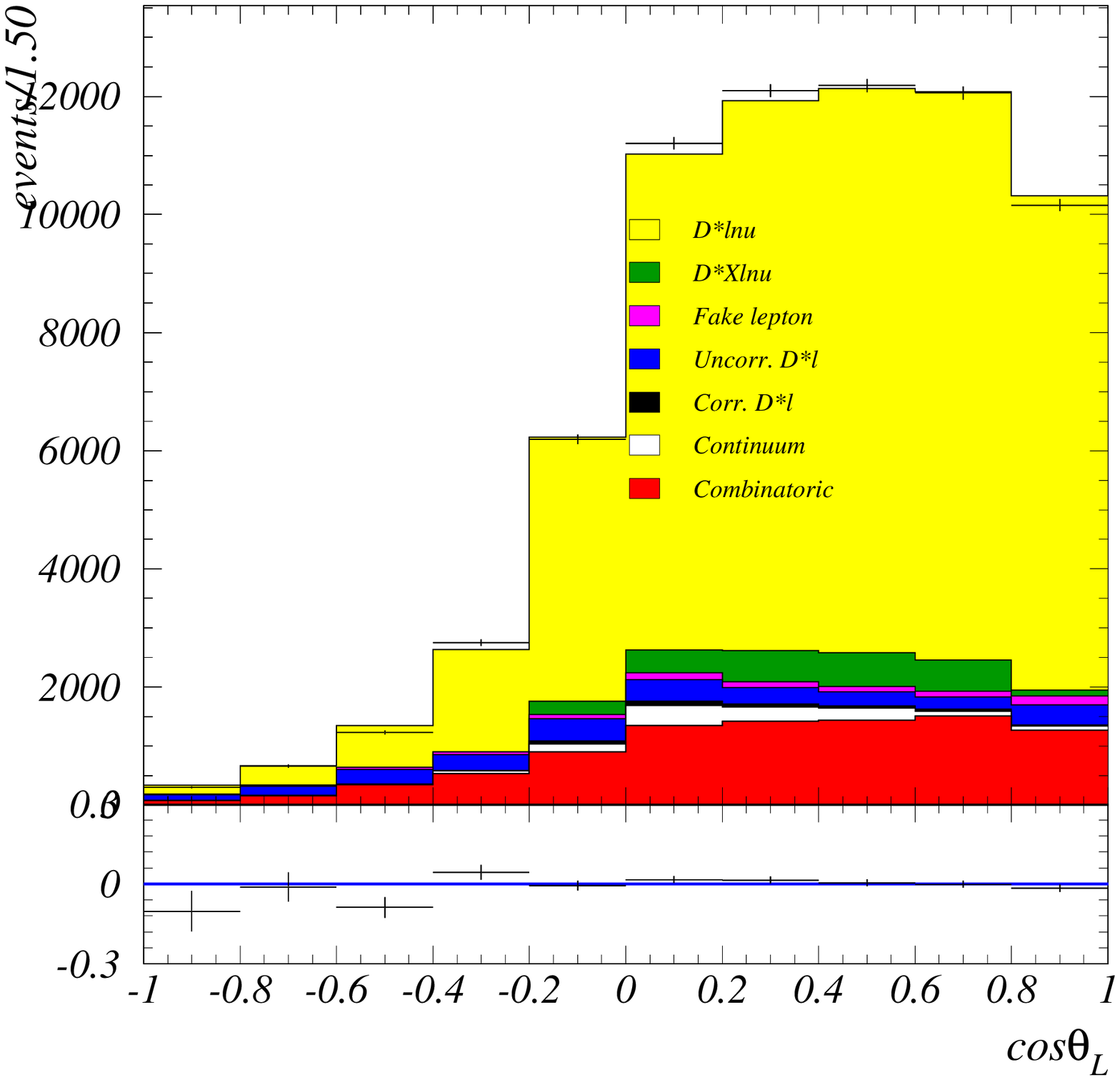}
\includegraphics[width=80mm]{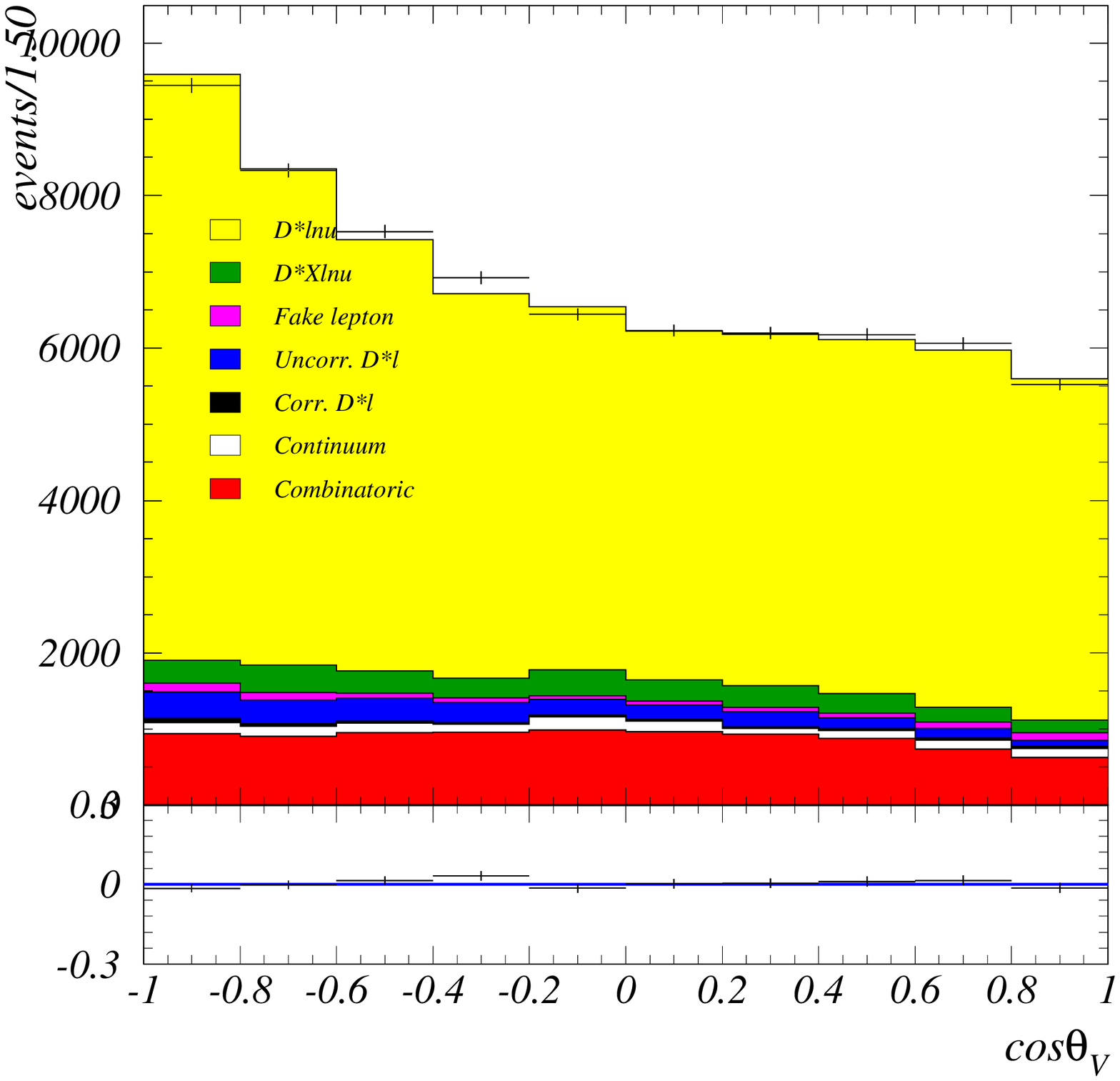}\includegraphics[width=80mm]{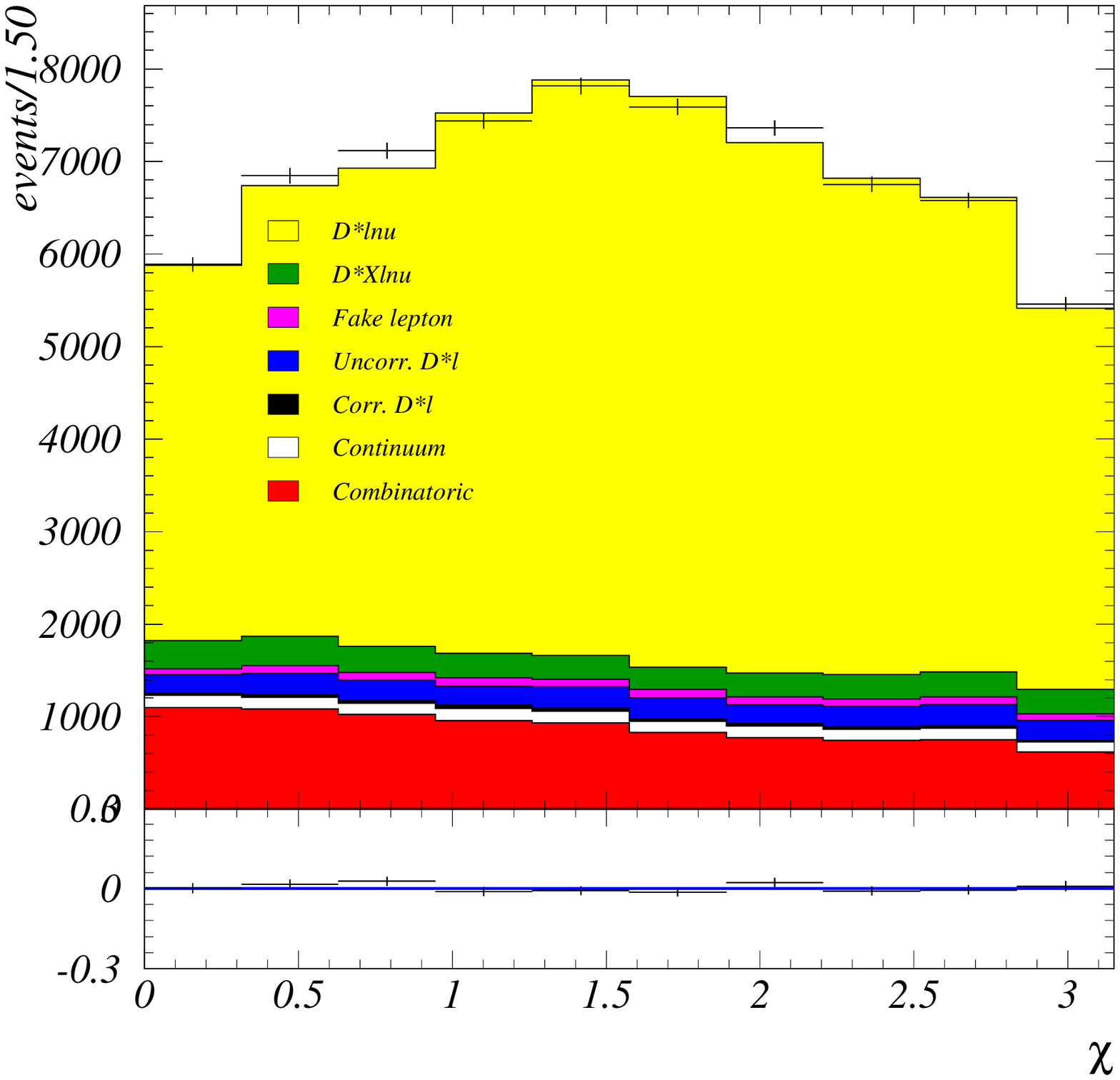}
\caption{\label{fig:DSexcl:projections}
Projections of data (dots) and simulation (histogram) to the four variables $w$, $\theta_V$, $\theta_\ell$ 
and $\chi$.}
\end{figure*}

Figure \ref{fig:DSexcl:projections} shows the projections of data an Monte Carlo to the three fitted variables,
as well as on the angle $\chi$, not used for the result.

Using the result from Lattice QCD calculation \cite{DSexcl:hashimoto-LQCD} $h_{A1}(1)=0.919^{+0.030}_{-0.035}$
as normalization and combining the measurement with an earlier result using electrons only \cite{DSexcl:babar-e},
\babar\ finds \cite{DSexcl:babar-comb}
\begin{eqnarray}
R_1 &=& 1.417\pm0.061\pm0.044 \nonumber \\
R_2 &=& 0.836\pm0.037\pm0.022 \nonumber \\
\rho^2 &=& 1.179\pm0.048\pm0.028 \nonumber \\
|V_{cb}| &=& (37.74 \pm 0.35_{stat} \pm 1.25_{syst} \,^{+1.23}_{-1.44} \,_{\Gamma_{SL}} ) \cdot 10^{-3} \nonumber
\end{eqnarray}
where the first uncertainty denotes the statistical precision from the fit, the second one gives the systematic
uncertainties and the third error quoted for the result for \VCB\ comes from the uncertainty of the total
semileptonic decay rate.

These results can be converted into a measurement of the branching fraction for neutral $B$ decaying semileptonicly
into charged \DS:
\begin{displaymath}
\BR(B^0 \to D^{*-}\ell^+\nu) = (4.77\pm0.04\pm0.39)\%.
\end{displaymath}

\subsection{Narrow Orbitally Excited Charm States $D^{**}$}

A sizable part of the background for the analysis discussed above, and also for many other analyses, comes
from semileptonic $B$ decays into higher excited charm states. \babar\ has presented preliminary results of
a measurement for the narrow resonances with $L=1$.

In Heavy Quark Symmetry (HQS) the spin of the heavy quark decouples, giving two possibilities for the sum of
the spin of the light quark, $s_q$, and the relative angular momentum, $L=1$, of $j_q=1/2$ or $j_q=3/2$.
With a finite mass for the heavy quark those configurations build two doublets, thus we have four
orbitally excited states.

The $j_q=1/2$ doublet, namely the $D^*_0$ and the $D^*_1$, can decay via S-wave transitions to the \D\ 
or \DS\ and a pion. Therefore these states are broad resonances.

\begin{figure}
\includegraphics[width=80mm]{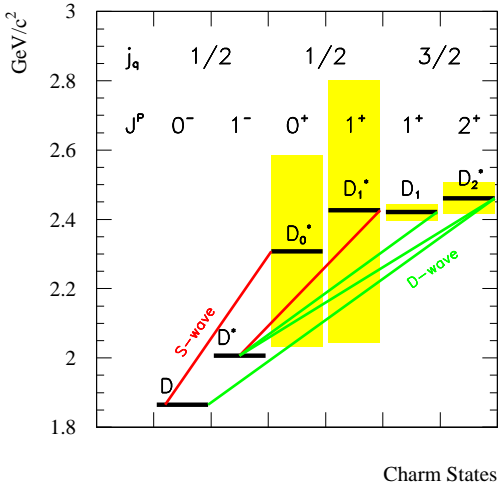}
\caption{\label{fig:DSS:states}
Masses and quantum numbers of the ground-states and orbitally excited states for charmed mesons.
The yellow bands indicate the measured widths of the states, the lines show the decays under pion
emission via S-wave (red) or D-wave (green).}
\end{figure}

For the $j_q=3/2$ doublet (\DI\ and \DII) parity conservation requires decays via D-wave transitions,
giving these resonances narrow widths of the order of few 10\mev. Conservation of angular momentum
restricts the \DI\ to decay into \DS\PI, while the \DII\ can decay into both, \D\PI\ and \DS\PI\
(see Figure \ref{fig:DSS:states}).

Violation of HQS allows a mixing of the $J^P=1^+$ states of the two doublets. This in principle allows
the \DI\ to decay via S-wave as well, but from the measured widths of the states one can conclude that
the S-wave contribution to the \DI\ decay is small.

\DSS\ candidates are reconstructed in four exclusive decay modes: $D^{*0} \pi^+$, $D^{*-} \pi^+$,
$D^0 \pi^+$ and $D^- \pi^+$. Then the candidates are paired with electrons or muons having a momentum
of $p_\ell>0.8\gevc$ in the center-of-mass frame. Again $\cos \theta_{B,D^{*(*)}\ell}$ provides the most
powerful tool to reduce backgrounds from other decays. It is used to select signal events, requiring
$|\cos \theta_{B,D^{**}\ell}|<1.$, as well as an explicit veto against combinatorial background coming from
true $B\to\DS\ell\nu$ decays by the requirement $\cos \theta_{B,D^*\ell}<-1$, taking only the \DS\ from the
reconstructed decay chain as hadronic part of the final state.

\begin{figure}[b]
\includegraphics[width=80mm]{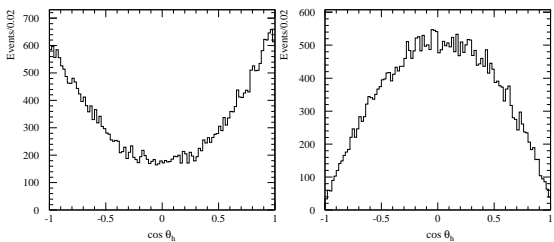}
\caption{\label{fig:DSS:helicity}
Distribution of the helicity angle $\cos \vartheta_h$ for \DS\ produced by \DI\ (left) and
\DII\ (right) taken from simulated signal events.}
\end{figure}

As pointed out above, the two decay modes $\DSS\to D\pi$ contain only signal from \DII\ while
the two modes $\DSS\to\DS\pi$ have an admixture of \DI\ and \DII. The mass difference of about 40\mevcc\
is too small to separate the two signals by mass information alone. A better handle gives the helicity
angle of the \DS\ in these modes. Depending on the total angular momentum of the mother \DSS, the
polarization of the daughter \DS\ differs. In case of the \DII\ the spin of the \DS\ must be aligned
in the direction of flight by angular momentum conservation. This gives a distribution of the helicity
angle $\cos \vartheta_h$ proportional to $\sin^2 \vartheta_h$. For the \DI\ all three polarizations can be
produced, leading in the sum to a distribution proportional to $A+\cos^2 \vartheta_h$.
Here the parameter $A$ depends on the (possible) polarization of the mother \DI\ and on the magnitude of
S-wave contributions to the decay caused by mixing of the two $J^P=1^+$ states. For an unpolarized
sample of \DI\ purely decaying via D-wave $A$ is expected to be $A=2$. Figure \ref{fig:DSS:helicity}
shows the expected helicity distributions for simulated signal events.

To make use of the helicity information, the two decay modes $\DS\pi$ are split up into 4 bins of
$|\cos \vartheta_h|$ each. Together with the two decay chains $\DSS\to D\pi$ this gives 10 modes.
All 10 distributions of $\Delta m=m(\DSS)-m(\DZ)$ are then fitted in common. Parameters of the fit are the
four branching ratios $\BR (B\to\DSS\ell\nu)$ (two types of \DSS\ with charge $\pm1$ or 0), masses and widths
of the states and finally the parameter $A$ describing the helicity distribution in the \DI\ decay and
the ratio $\Gamma(\DII\TO\DS\PI)/\Gamma(\DII\TO\D\PI)$ which has not been determined yet.

\begin{figure*}
\includegraphics[width=160mm]{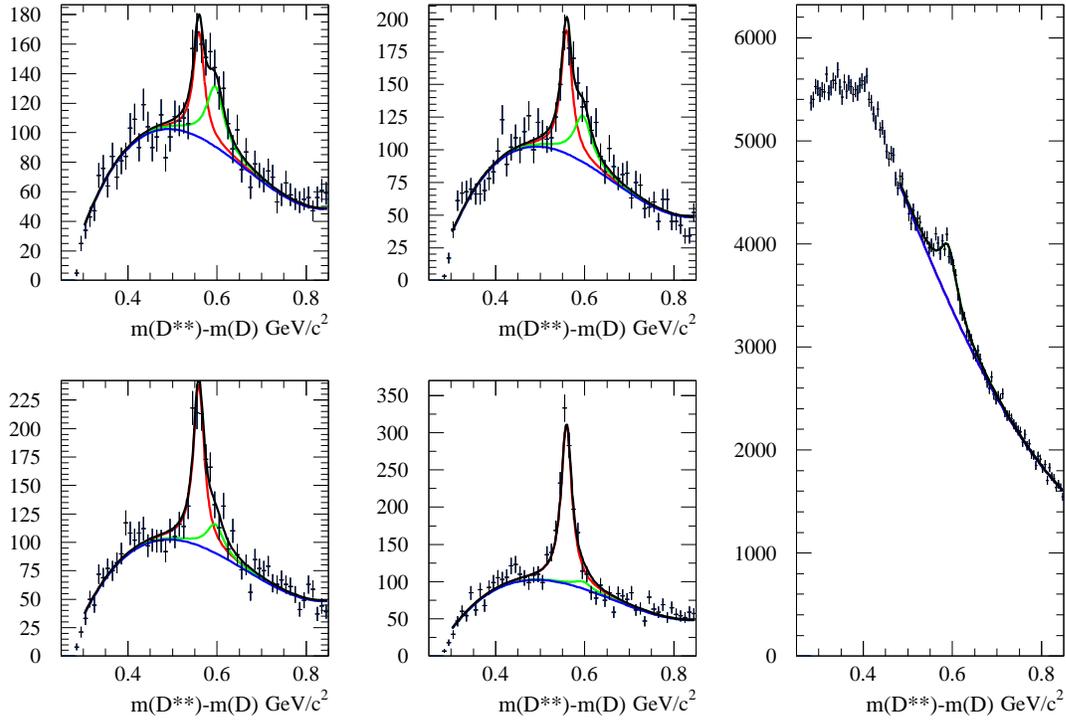}
\caption{\label{fig:DSS:fit}
Spectra of $\Delta m$ for neutral \DSS\ candidates (data) in the mode \D\PI\ (right) and \DS\PI\
(left) for the four different bins in helicity ($|\cos \vartheta_h|<0.25$, $0.25<|\cos \vartheta_h|<0.5$, 
$0.5<|\cos \vartheta_h|<0.75$, and $0.75<|\cos \vartheta_h|$ from upper left to lower right).
Superimposed are the fit result (black) and its contributions from \DI\ (red), \DII\ (green), and
backgrounds (blue).}
\end{figure*}

Figure \ref{fig:DSS:fit} shows the spectra of $\Delta m$ for neutral \DSS\ candidates together with the
fitted contributions.

For the decays of the \DSS\ Isospin is assumed to be conserved. In addition, the modes $\DSS\to D^{(*)}\pi$
are assumed to saturate the \DSS\ decays. Analyzing 208\fb\ of data \babar\ has reported preliminary results
to be
\begin{eqnarray}
\BR (B^+ \to D_1^{0}  \ell^+ \nu_{\ell}) &=& ( 4.48 \pm 0.26 \pm 0.35) 10^{-3} \nonumber \\
\BR (B^+ \to D_2^{*0} \ell^+ \nu_{\ell}) &=& ( 3.54 \pm 0.32 \pm 0.54) 10^{-3} \nonumber \\
\BR (B^0 \to D_1^{-}  \ell^+ \nu_{\ell}) &=& ( 3.64 \pm 0.32 \pm 0.49) 10^{-3} \nonumber \\
\BR (B^0 \to D_2^{*-} \ell^+ \nu_{\ell}) &=& ( 2.70 \pm 0.35 \pm 0.43) 10^{-3} \nonumber
\end{eqnarray}
where the first uncertainty reflects the statistical precision of the fit and the second one
denotes systematic effects.

An interesting note on these preliminary results is, that the ratio between the production of
\DII\ and \DI\ comes out to be about $\BR_{\DII}/\BR_{\DI} \approx 0.8$. This parameter is important
to distinguish between different models in HQET. The \babar\ numbers are contrary to results
reported by the $\mbox{D}\emptyset$ Collaboration, that finds, although with large uncertainties,
a ratio larger than 1. \cite{DSS:D0.ratio}.

Recently, Belle has reported the observation of decays $\DSS\to D\pi\pi$ \cite{DSS:belle.D1-nonres}.
The ratio of these decays to the resonant ones $\DS\pi$ can be deduced from the measurements of hadronic
$B$ decays to be about 20\%. Taking this into account, the numbers reported by \babar\ need to be
scaled by that factor.

\subsection{Combined Measurement of Exclusive Branching Fractions}

Also aiming to a better understanding of the exclusive contributions to the inclusive semileptonic
width is another analysis recently published by \babar. Contrary to the analyses discussed before,
no particular exclusive semileptonic decay mode is reconstructed apart from a $D$ candidate and a lepton.
Instead the decay of the other $B$ of the event ($B_{tag}$) is reconstructed in hadronic modes and the
particular type of the semileptonic decay of the signal side is deduced from all particles not being used already.

Electrons and muons are required to have a momentum of $p_\ell>0.6\gevc$ in the center-of-mass frame.
Neutral $D$ mesons are reconstructed in 9 different channels: $K (1..4)\pi$, $KK$ and $\pi\pi$
containing not more than one \PIZ. Charged $D$ mesons are similarly reconstructed in 7 modes: $K (1..4)\pi$,
$K^-K^+\pi$ and $K^0_S K$, again with no more than one \PIZ\ included.

One of those $D$ candidates is now combined with further particles to form a decay $B\to D^{(*)} Y$, where 
$Y$ consists of charged and neutral pions and kaons of the kind
$Y=n_{\PIC} \PIC + n_{K^\pm} K^{\pm} + n_{\PIZ} \PIZ + n_{K^0_s} K^0_s$. The total charge of $Y$ is required
to be $\pm1$ and for the number of particles the requirements are $n_{\PIC} + n_{K^\pm}\le 5$ and
$n_{\PIZ},n_{K^0_s}\le 2$. In total this gives a variety of about 1000 different decay modes to reconstruct
hadronic $B$ decays.
A second $D$ candidate then is taken as part of the hadronic final state of the semileptonic decay of the
signal $B$. 

To distinguish between semileptonic decays to $D\ell\nu$, $\DS\ell\nu$ and $\DSS\ell\nu$, the last being
a generic term for the orbitally excited states discussed above as well as for other excited charm states
and non-resonant decays $D^{(*)}\pi\ell\nu$, three quantities are used:
From the rest of the event, that are all charged and neutral particles not used to reconstruct the tag $B$ or
the $D\ell$ for the signal, the invariant mass $m^2_{miss}$ and the number of charged tracks $N_{trk}$, and
finally the lepton momentum $p_\ell$.

\begin{figure*}
\includegraphics[width=160mm]{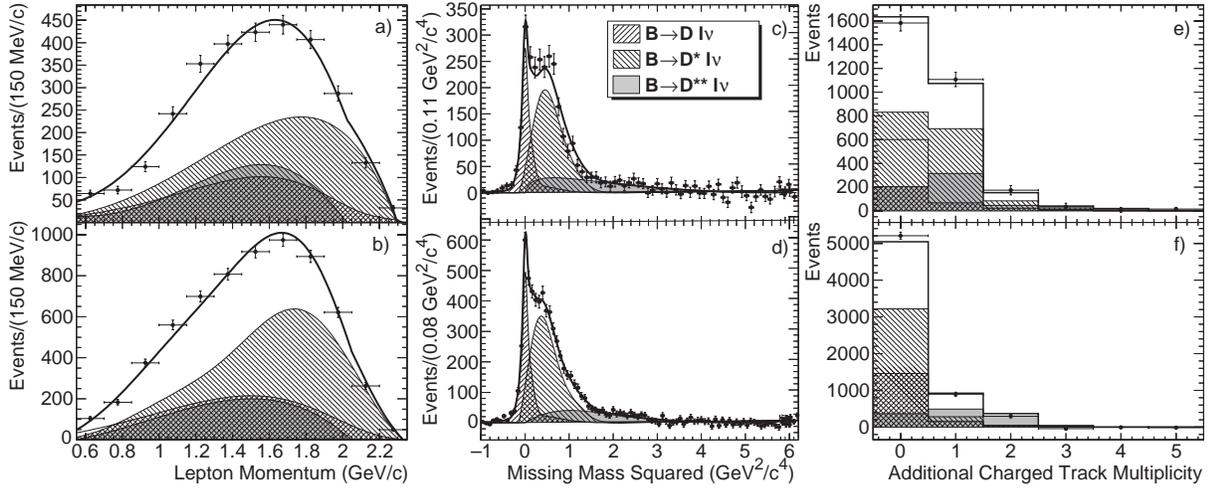}
\caption{\label{fig:combBR:fit}
Distributions of the three quantities $p_\ell$, $m^2_{miss}$ , and $N_{trk}$ for semileptonic decays of
$B^0$ (top) and $B^\pm$ (bottom). Points are data, the solid line gives the fit result and the shaded
functions indicate the contributions from the three different decay modes.}
\end{figure*}

To describe the distributions in these three variables, probability density functions (PDF) are build from
data in order to reduce systematic uncertainties coming from simulations. For all the three decay modes
exclusive selections are defined, mainly based on $m^2_{miss}$ and the mass difference $\Delta m = m(D\pi)-m(D)$
which provides a clean signature for \DS. The purity reached by the exclusive selections are in the range of
75-91\% depending on the decay mode in question. Since $m^2_{miss}$, $N_{trk}$, and $p_\ell$ are largely
uncorrelated, the distributions deduced from the exclusively selected subsets can be used as PDF's for the
inclusive data set. Remaining correlations are studied and treated as systematic uncertainty.

The fit is performed on 340\fb\ of data and illustrated in Figure \ref{fig:combBR:fit}.
The total $\chi^2$ per degree of freedom is 1.21 and 0.94 for neutral and charged $B$ respectively.
Table \ref{tab:combBR:result} summarizes the results \cite{CombBR:babar}.

\begin{table}[h]
\begin{center}
\caption{\label{tab:combBR:result}
Results of the combined fit to exclusive branching fractions for charged and neutral $B$ decays. The
first uncertainty denotes the statistical precision from the fit, the second one systematic effects.}
\begin{tabular}{|c|c|c|}
\hline
Ratio & $B^-$ (\%) & $B^0$ (\%) \\ \hline
{\large $\frac{\Gamma(B\to D \ell\nu)}{\Gamma(B\to DX \ell\nu)}$} &
$22.7\pm1.4\pm1.6$ & $21.5\pm1.6\pm1.3$ \\ \hline
{\large $\frac{\Gamma(B\to D^* \ell\nu)}{\Gamma(B\to DX \ell\nu)}$} &
$58.2\pm1.8\pm3.0$ & $53.7\pm3.1\pm3.6$ \\ \hline
{\large $\frac{\Gamma(B\to D^{**} \ell\nu)}{\Gamma(B\to DX \ell\nu)}$} &
$19.1\pm1.3\pm1.9$ & $24.8\pm3.2\pm3.0$ \\ \hline
\end{tabular}
\end{center}
\end{table}

So far, no measurements of the total rate $\Gamma(B\to DX \ell\nu)$ is available. However, a good approximation
is given by $\Gamma(B\to DX \ell\nu) \approx \Gamma(B\to X_c \ell\nu)$. Possible charm states $X_c$ that do not
cascade down to $D X$ are $D_s$ mesons and charmed baryons. Those are expected to have branching fractions of
the order of $\mathcal{O}(10^{-4})$ and less justifying the assumption made.

Taking the measurement for $\Gamma(B\to X_c \ell\nu)$ as normalization, one finds the following branching
fractions:
\begin{eqnarray}
\BR(B^-\to D^0 \ell^- \nu)     &=& ( 2.42 \pm 0.15 \pm 0.17 ) \% \nonumber \\
\BR(B^-\to D^{*0} \ell^- \nu)  &=& ( 6.20 \pm 0.19 \pm 0.32 ) \% \nonumber \\
\BR(B^-\to D^{**0} \ell^- \nu) &=& ( 2.04 \pm 0.14 \pm 0.20 ) \% \nonumber \\
\BR(B^0\to D^+ \ell^- \nu)     &=& ( 2.19 \pm 0.16 \pm 0.13 ) \% \nonumber \\
\BR(B^0\to D^{*+} \ell^- \nu)  &=& ( 5.46 \pm 0.33 \pm 0.37 ) \% \nonumber \\
\BR(B^0\to D^{**+} \ell^- \nu) &=& ( 2.52 \pm 0.32 \pm 0.31 ) \%. \nonumber
\end{eqnarray}
These results are in precision comparable to the current world knowledge \cite{pdg2006}.

\section{Inclusive Decays and Analysis of Moments}

Contrary to the approach to extract \VCB\ from a single decay mode are inclusive measurements.
These analyses measure the total width $\Gamma_{c\ell\nu}$ which is proportional to \VCB\ and
corrections to the first order result:
\begin{displaymath}
\Gamma_{c\ell\nu}=\frac{G_F m_b^5}{192 \pi^3} |V_{cb}|^2 (1+A_{ew})A_{pert}A_{nonpert}.
\end{displaymath}
Here $A_{ew}$ are corrections from electroweak interaction which are theoretically well under control.
$A_{pert}$ and $ A_{nonpert}$ are corrections arising from QCD and can be grouped in first a
perturbative expansion in $\Lambda_{QCD}/m_b$ giving $A_{pert}$, and second non-perturbative
corrections in powers of $1/m_b$, $A_{nonpert}$.

Calculations for those QCD corrections are done in the framework of Operator Product Expansion (OPE).
The two most prominent approaches are calculations in the so called 'kinetic scheme' \cite{HQE:kin-scheme}
and the '1S scheme' \cite{HQE:1s-scheme}. They differ in the choice of the mass scale and hence have
different sets of input parameters. The first order parameters are the quark masses $m_b$ and $m_c$ and
two expectation values: The one of the kinetic operator (called $\mu_\pi^2$ in the kinetic scheme or
$\lambda_1$ in the 1S scheme) which describes the motion of the heavy quark inside the hadron and the 
one of the chromomagnetic operator (called $\mu_G^2$ in the kinetic, $\lambda_2$ in the 1S scheme).

In order to get results for \VCB, the complete set of input parameters $\vec a$ need to be determined.
Therefore a measurement of the semileptonic width $\Gamma_{c\ell\nu} \cong |V_{cb}|^2 f_{OPE}(m_b, m_c,\vec a)$
is not sufficient. Instead several kinematic quantities in semileptonic $B$ decays are measured, their
shape is characterized by the moments of their distributions. Those moments now can also be calculated
within OPE in terms of the theory parameters $m_b, m_c,$ and $\vec a$ and a fit of the measured moments
to the theoretical predictions determines the theory parameters as well as \VCB.

If the number of measurements used over-constrains the fit this technique also gives sensitivity to
the validity of the underlying assumptions of the calculation.

\subsection{\VCB\ and HQE Parameters}

The most recent analysis of kinematic moments in inclusive decays is reported by
the Belle Collaboration, making use of 140\fb\ of data. First one $B$ is reconstructed exclusively in
a large variety of hadronic decay modes with high purity. Then on the signal side an electron is reconstructed
as signature for a semileptonic decay and the rest of the event is treated as the hadronic final state.
Backgrounds from non-\BB\ events and $b\to u \ell\nu$ transitions are subtracted to provide an inclusive
sample of decays $B\to X_c\ell\nu$.

From this sample moments are deduced for the electron energy $E_e$ and the hadronic mass spectrum $M_X^2$.
Since it is impossible to cleanly identify electrons down to infinite small energies a cut-off value need
to be placed. This minimal allowed energy is varied over a large range and the moments are determined
as functions of this cut on $E_e$.

\begin{figure}
\includegraphics[width=42mm]{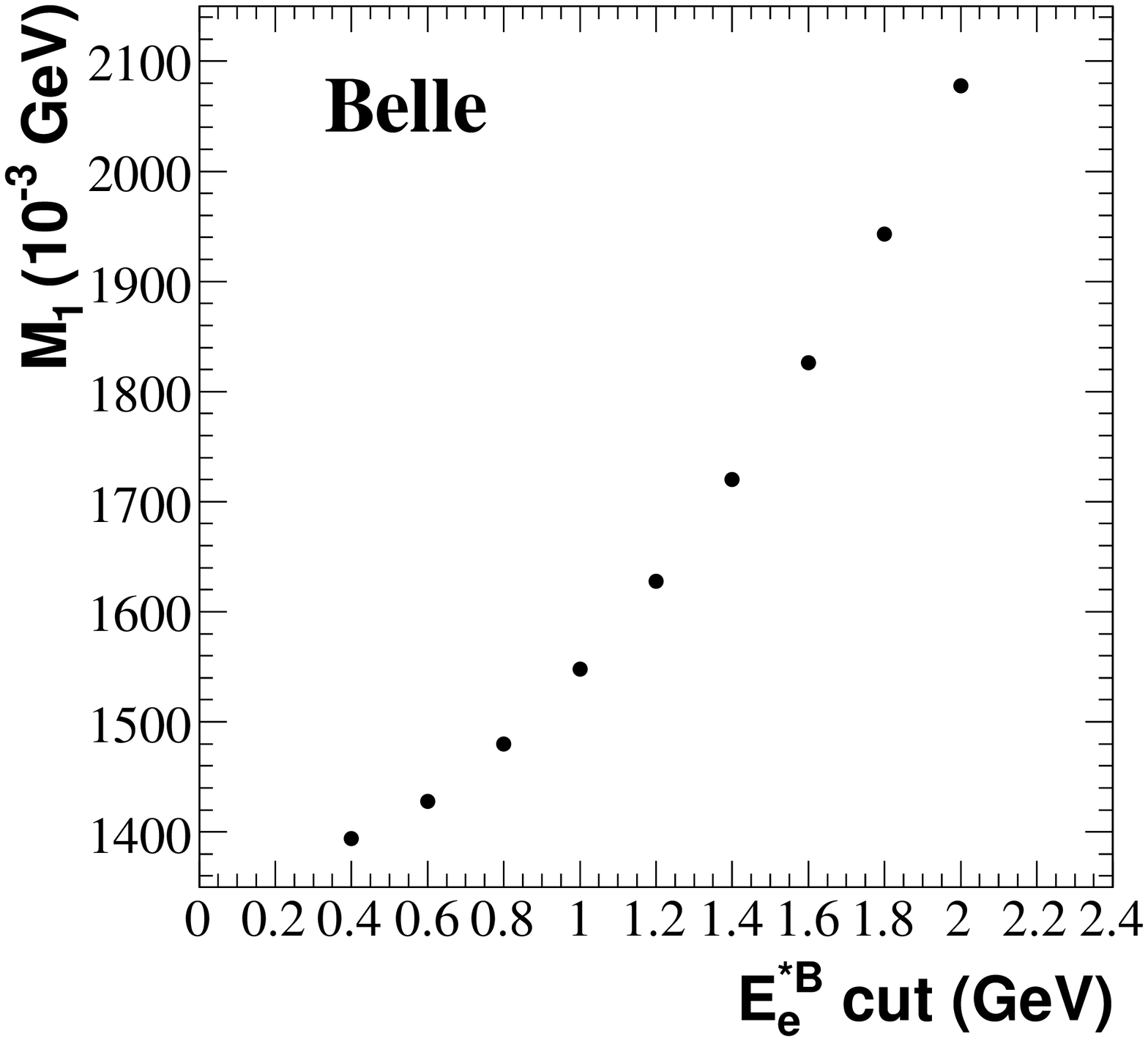}\includegraphics[width=42mm]{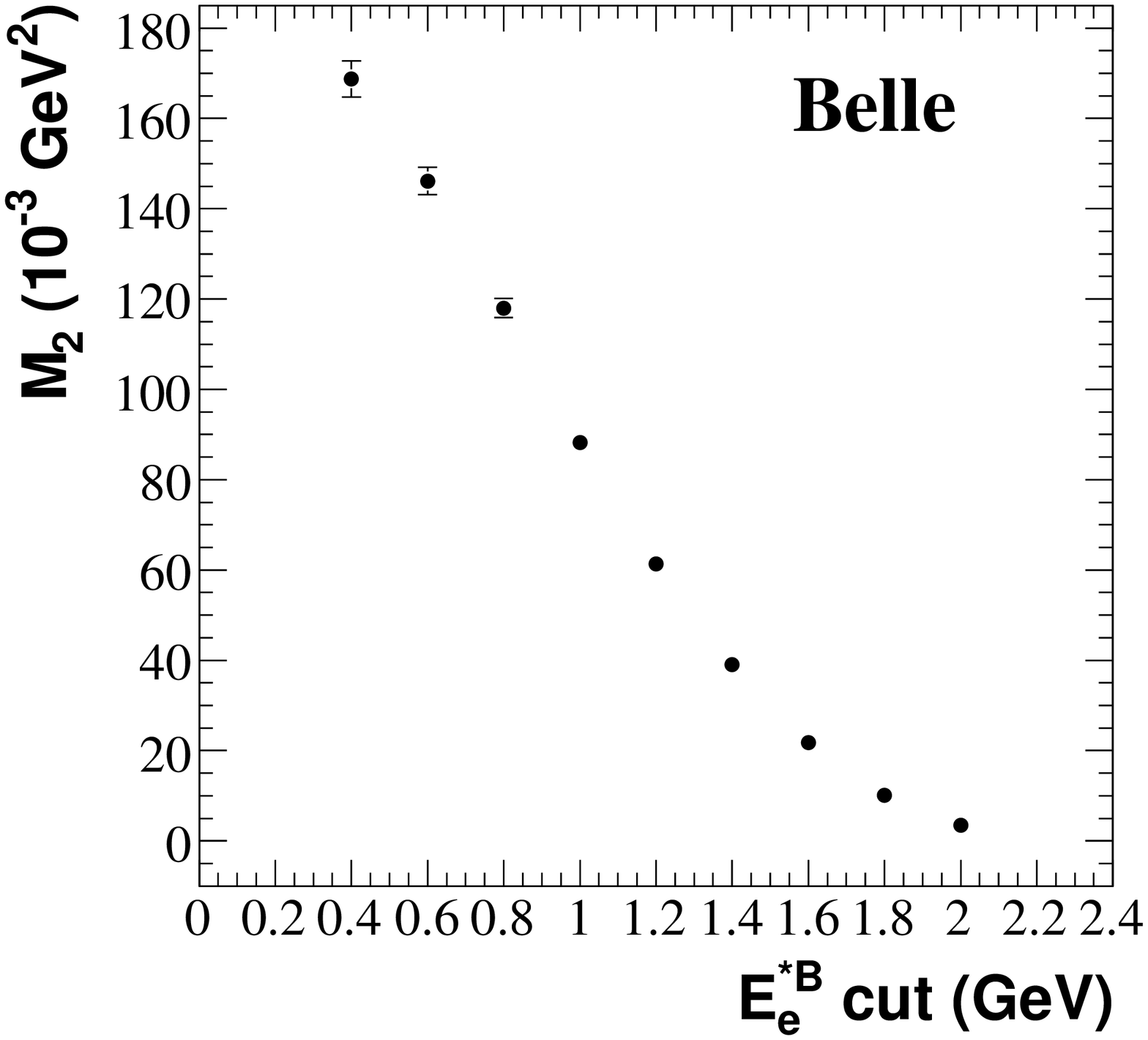}
\includegraphics[width=42mm]{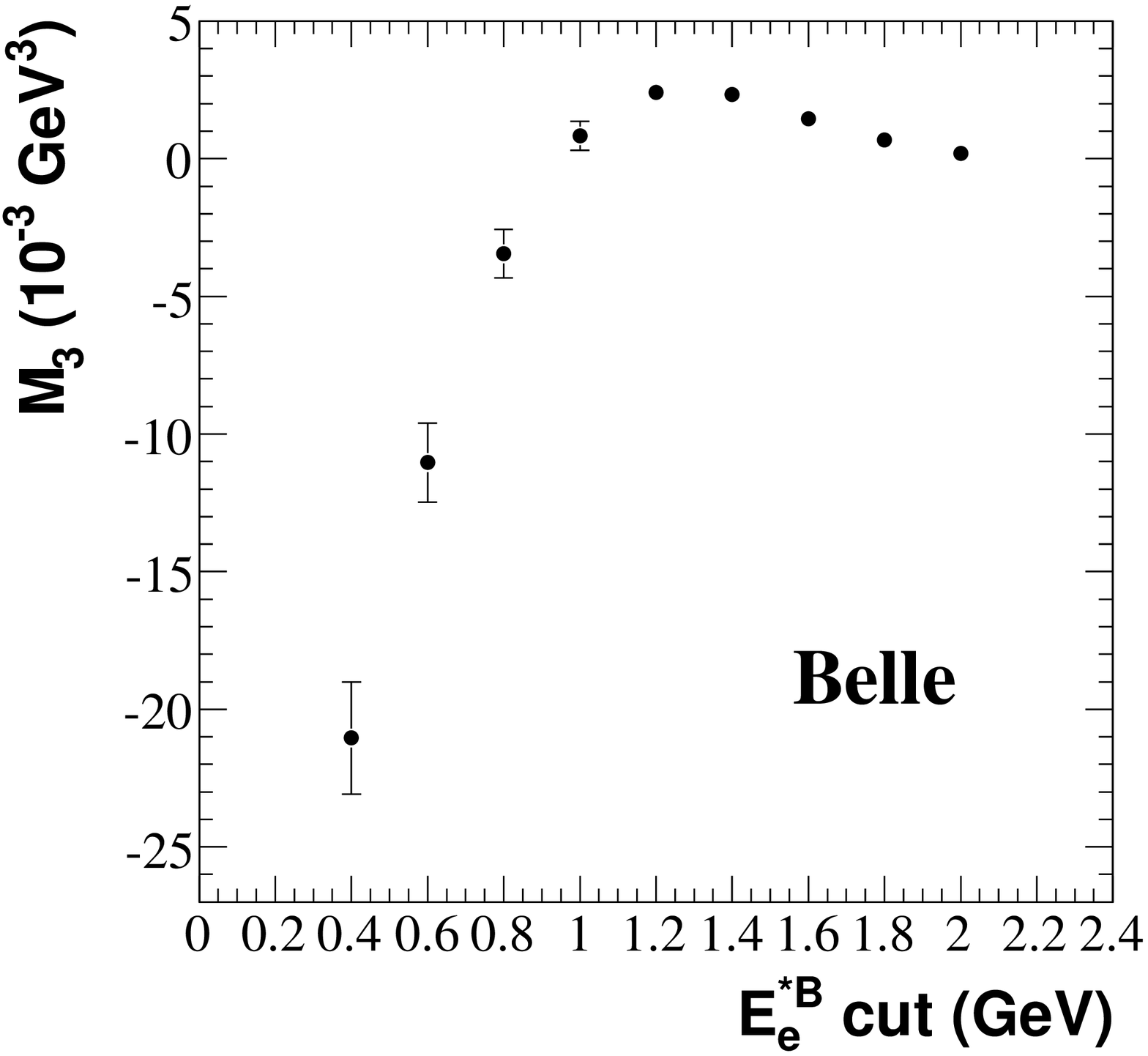}\includegraphics[width=42mm]{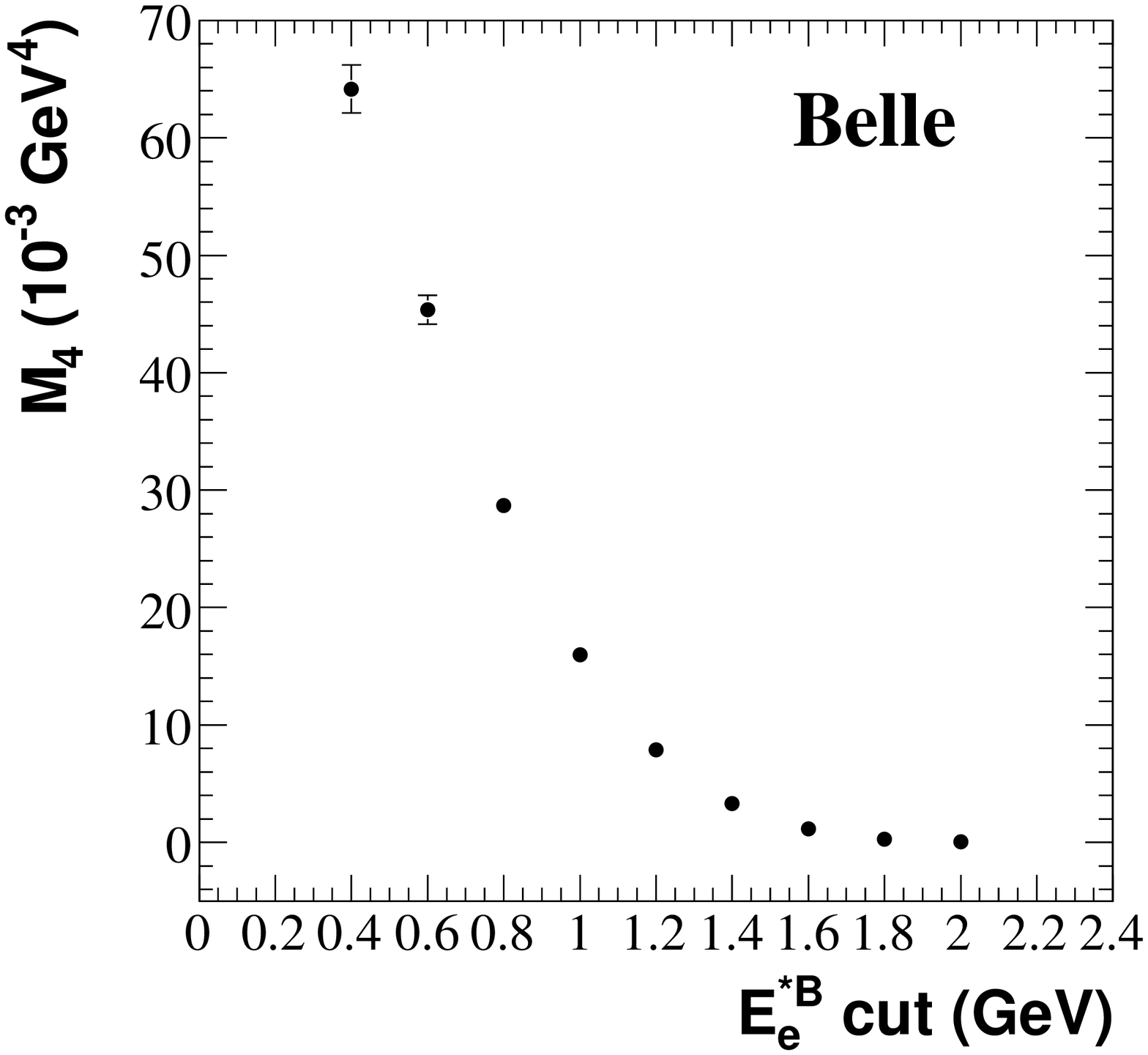}
\caption{\label{fig:belle-ee}
Distributions of the first four moments (from top left to bottom right) of
the electron energy spectrum as function of the minimal allowed electron energy.}
\end{figure}

The electron energy spectrum is distorted by detector effects and the true energy is unfolded from the
measured distribution assuming the electron to be massless. For the electron energy $E_e$ the cut-off
value is varied between 0.4 and 2.0\gev\ in the $B$ rest frame. From the spectrum the first four moments
are calculated taking all correlations between the different cut-off values into account \cite{HQE:belle-ee}.
Figure \ref{fig:belle-ee} shows the results.

The total number of events can be translated into the semileptonic branching fraction restricted to the
electron energy range in question. For the lowest cut-off Belle reports a result of
\begin{eqnarray}
\BR(B^+)|_{E_\ell>0.4\mbox{\scriptsize GeV}}&=&(10.79\pm0.25\pm0.27)\cdot 10^{-3} \nonumber \\
\BR(B^0)|_{E_\ell>0.4\mbox{\scriptsize GeV}}&=&(10.08\pm0.30\pm0.22)\cdot 10^{-3}. \nonumber
\end{eqnarray}

\begin{figure}
\includegraphics[width=80mm]{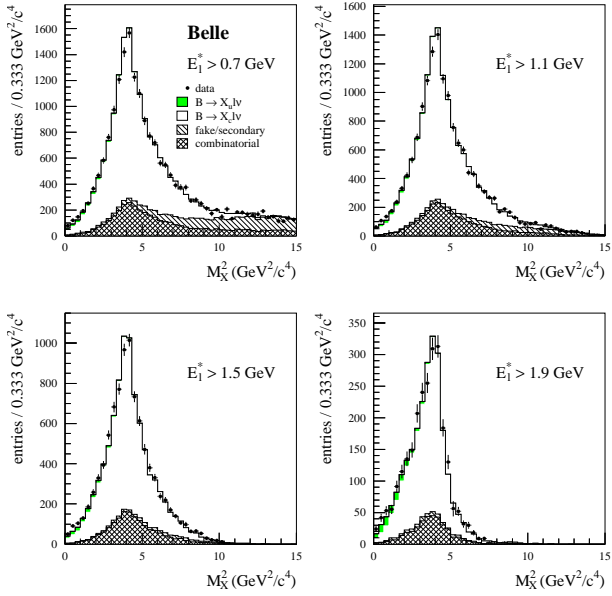}
\caption{\label{fig:belle-mhad}
Unfolded distribution of the square of the hadronic mass $M_X^2$ for four different values
of the lepton energy cut-off. Dots with error bars represent the data and histograms are
simulations of signal and various background sources.}
\end{figure}

\begin{figure}
\includegraphics[width=80mm]{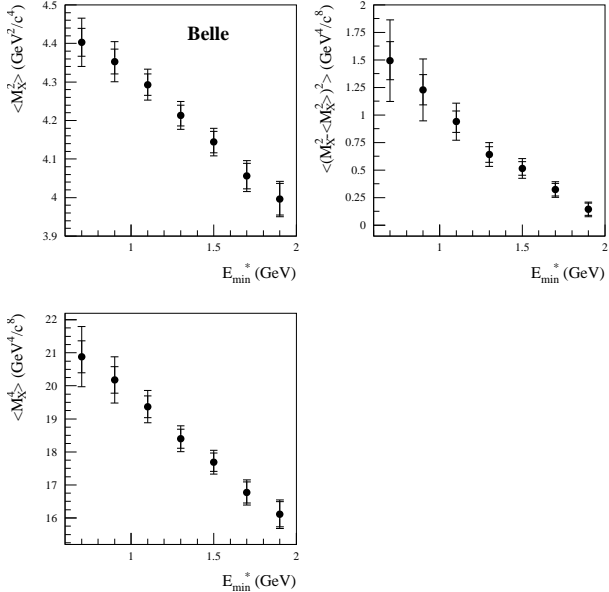}
\caption{\label{fig:belle-m2}
Distribution of the first (top left), second central (top right) and second non-central moment
(bottom left) of the hadronic mass $M_X^2$ as functions of the minimal allowed lepton energy.}
\end{figure}

For the hadronic mass distribution electrons and muons are used for the lepton and the minimal lepton
energy is varied between 0.7 and 1.9\gev. To improve the resolution the hadronic mass is not
reconstructed from the hadronic final state of the semileptonic decay, instead it is calculated from
the beam energies, the reconstructed four momentum of $B_{tag}$, the measured lepton and the four
momentum of the neutrino which is set to equal the momentum of the missing mass of the event. Again
the true mass distribution is unfolded from the measured one and three moments of $M_X^2$ are
calculated: the first, the second central and the second non-central \cite{HQE:belle-mhad}.
Figure \ref{fig:belle-mhad} shows examples for the distribution of the missing mass for four different
values of the lepton energy cut-off and figure \ref{fig:belle-m2} shows the three moments as function of
this cut-off.

Taking the moments of $E_e$ and $M_X$ as described above from \cite{HQE:belle-ee} and 
\cite{HQE:belle-mhad} Belle extracts the HQE parameters and \VCB. Additional information
is used from decays $B\to X_s\gamma$ where the energy distribution of the photon is
connected to the motion of the $b$ quark inside the $B$ hadron. Belle has previously measured
moments of the photon energy $E_\gamma$ as well \cite{HQE:belle-egamma} and uses these
results as further input.

All together Belle fits a total of 71 truncated moments which are theoretically described
up to order $\mathcal{O}(1/m_b^3)$. This fit is performed in both OPE approaches, calculations in
the kinetic scheme as well as in the 1S scheme \cite{HQE:belle-params}.

\begin{figure}
\includegraphics[width=40mm]{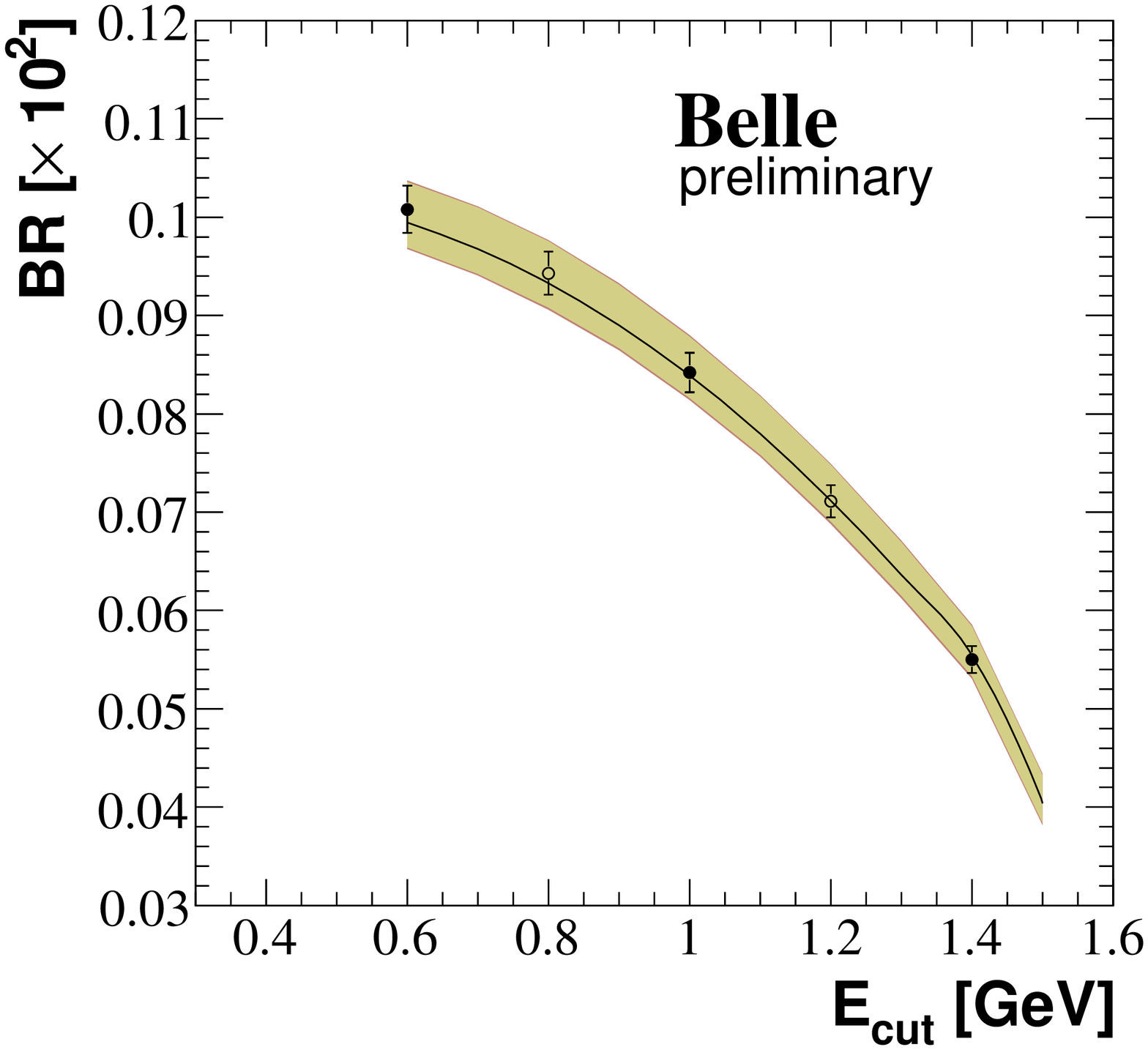}
\includegraphics[width=40mm]{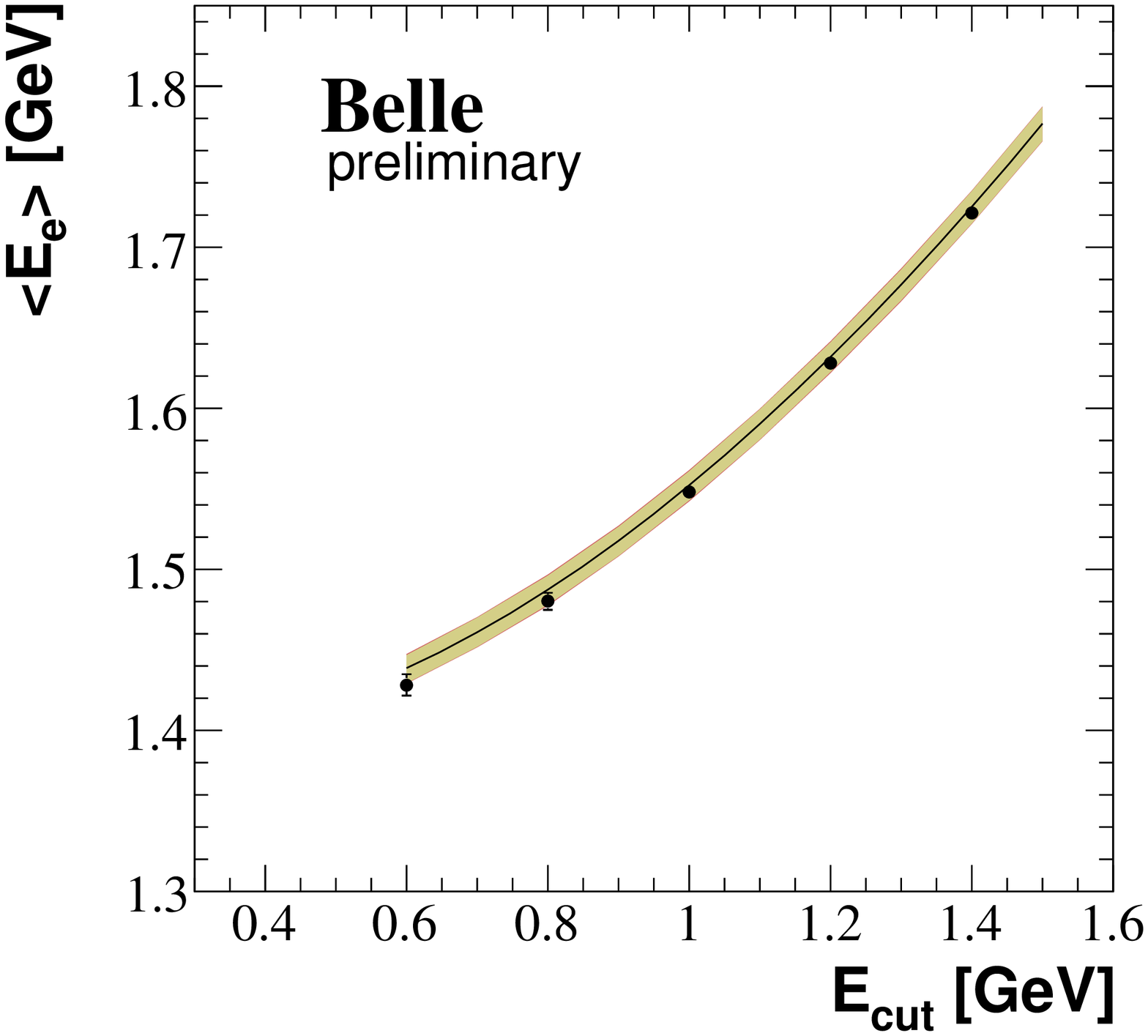}
\includegraphics[width=40mm]{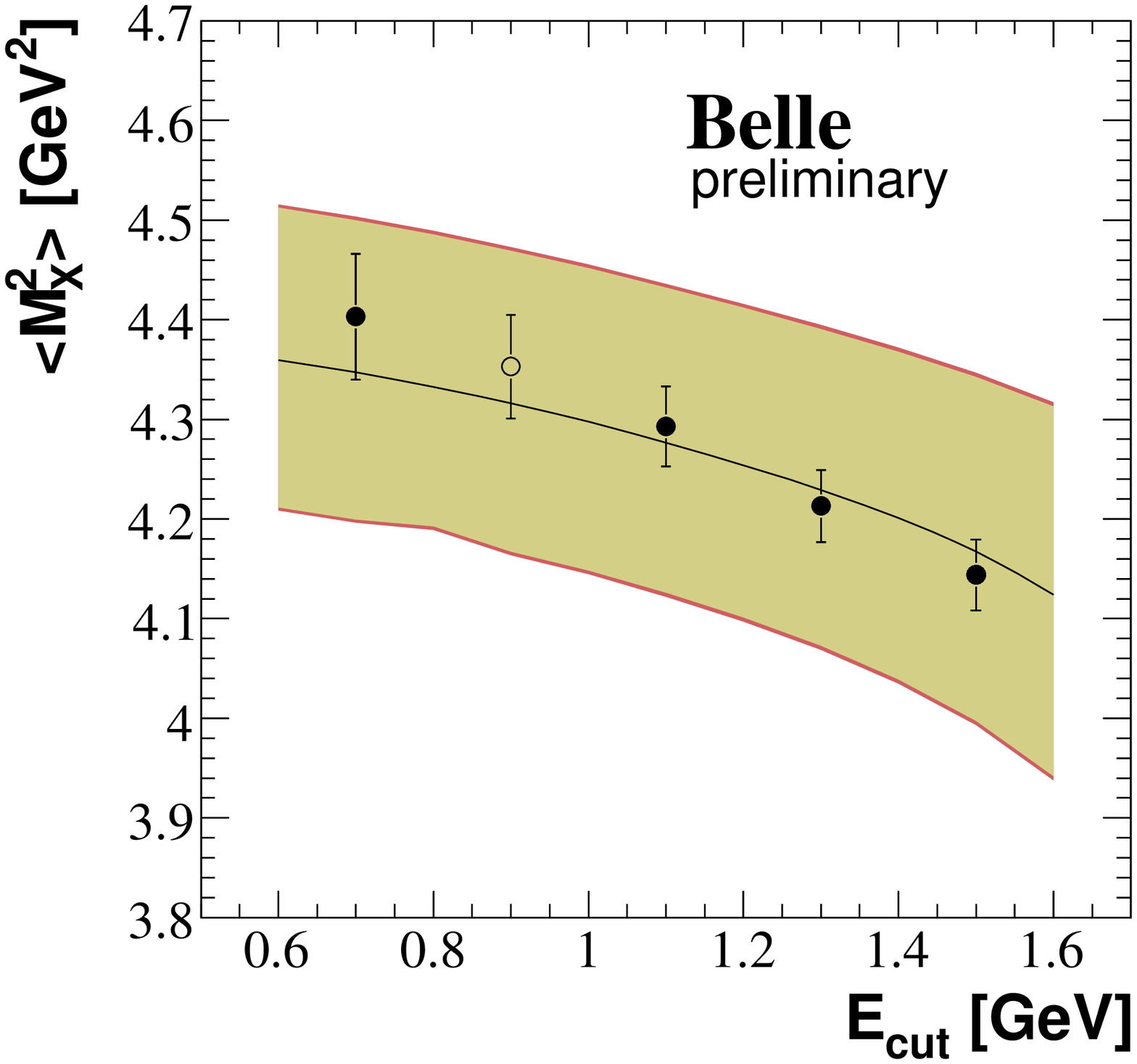}
\includegraphics[width=40mm]{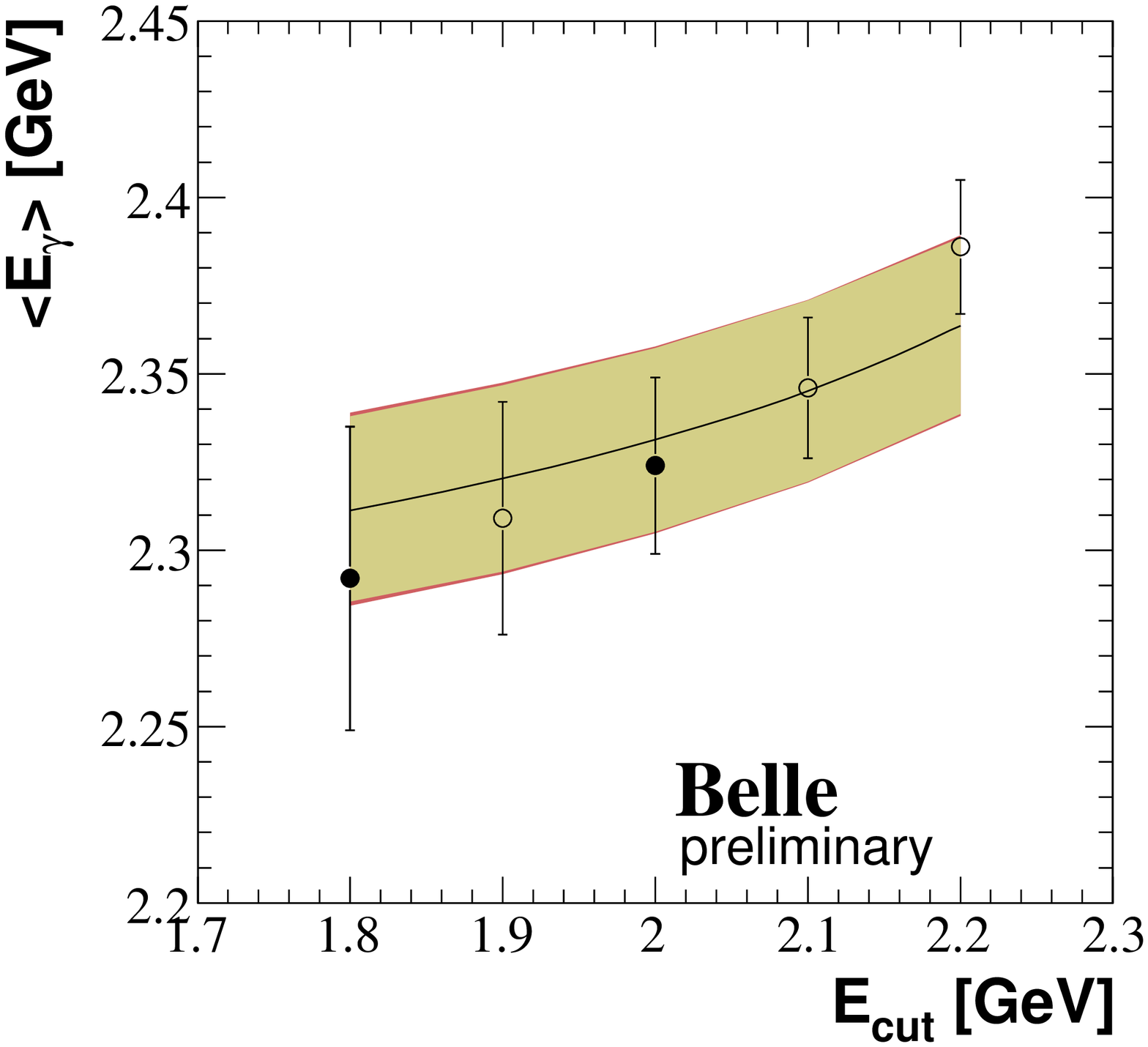}
\caption{\label{fig:HQEfit}
Selected results of the fit of 71 truncated moments in $E_e$, $M_X$ and $E_\gamma$ to the
predictions in the 1S scheme. Shown are, as function of the minimal allowed lepton energy, the
semileptonic branching fraction (top left), the first moment of the lepton energy (top right),
the first moment of the square of the hadronic mass (bottom left) and the first moment of the
photon energy (bottom right). Dots with error bars are the measurements where open circles
represent data points that are not used for the fit. The colored bands give the uncertainty from 
the fit (yellow) and the total uncertainty (red).}
\end{figure}

In the kinetic scheme a total of seven theory parameters is fitted together with \VCB\ as an
eights parameter. In the 1S scheme restrictions to the full parameter space needed to be
set. Thus some variables were expressed in terms of others or fixed to certain values, leaving
three independent fit parameters. Figure \ref{fig:HQEfit} gives as example the fit in the 1S scheme
for the total semileptonic branching fraction and the first moments of $E_e$, $M_X^2$ and $E_\gamma$.

In the kinetic scheme, Belle reports the following results for \VCB\ and the theory
parameters:
\begin{eqnarray}
|V_{cb}| &=& (41.93\pm0.65_{fit}\pm0.07_{\alpha_s}\pm0.63_{\Gamma_{SL}}) \cdot 10^{-3} \nonumber \\
\BR_{c\ell\nu} &=& ( 10.590 \pm 0.164_{fit} \pm 0.006_{\alpha_s} ) \% \nonumber \\
m_b      &=& ( 4.564 \pm 0.076_{fit}\pm 0.003_{\alpha_s} ) \gev \nonumber \\
m_c      &=& ( 1.105 \pm 0.116_{fit}\pm 0.005_{\alpha_s} ) \gev \nonumber \\
\mu^2_\pi &=& ( 0.557 \pm 0.091_{fit}\pm 0.013_{\alpha_s} ) \gev^2 \nonumber \\
\mu^2_G &=& ( 0.358 \pm 0.060_{fit}\pm 0.003_{\alpha_s} ) \gev^2 \nonumber \\
\tilde{\rho}^3_D &=& ( 0.162 \pm 0.053_{fit}\pm 0.008_{\alpha_s} ) \gev^3 \nonumber \\
\rho^3_{LS} &=& ( -0.174 \pm 0.098_{fit}\pm 0.003_{\alpha_s} ) \gev^3. \nonumber
\end{eqnarray}
Doing the fit, theoretical uncertainties of the calculated moments are included to the
calculation of the $\chi^2$. Thus the uncertainties of the results labeled as {\em fit}
represent both, the statistical precision and most of the theoretical uncertainties.
Only the uncertainty in the knowledge of $\alpha_s$ is treated separately and stated as
second error. For the result of \VCB\ the third error reflects the uncertainty on the
total semileptonic width $\Gamma_{SL}$.

In the 1S scheme, Belle finds the following results:
\begin{eqnarray}
|V_{cb}| &=& (41.49\pm0.52_{fit}\pm0.20_{\tau_B}) \cdot 10^{-3} \nonumber \\
m_b^{1S} &=& ( 4.729 \pm 0.048 ) \gev \nonumber\\
\lambda_1 &=& ( -0.30 \pm 0.04 ) \gev^2. \nonumber
\end{eqnarray}
Again the stated uncertainty is a combination of statistical precision and theoretical 
uncertainties. For the result of \VCB\ the influence of the lifetime of the $B$, $\tau_B$,
is given separately.

These results can be compared to previous measurements published by several other experiments
for the moments of $E_e$ \cite{HQE:other-ee} and $M_X$ \cite{HQE:other-mhad}, and the
subsequent extraction of the HQE parameters and \VCB\ by Oliver Buchm\"uller and Henning
Fl\"acher \cite{HQE:buchmueller-params}. Both results are in good agreement and have
comparable total error budgets.

\section{Summary}

There have been many improvements recently in our knowledge of semileptonic $B$ decays into
charmed mesons. Two contrary approaches, exclusive and inclusive measurements, allow to understand
the total semileptonic decay rate $\Gamma_{c\ell\nu}$ and its composition by exclusive decay
modes. The results from different experiments, datasets and analysis methods converge nicely
to a complete picture.

Using the full available datasets at the $B$ Factories will further improve the situation,
especially about exclusive decay modes. One of the main issues here are broad charm resonances
and non-resonant decays. Those should become accessible now using developed techniques to
reconstruct the other $B$. Also other open questions, as for example the conservation of Isospin,
might be answered with measurements of higher precision soon.

\begin{figure}
\includegraphics[width=80mm]{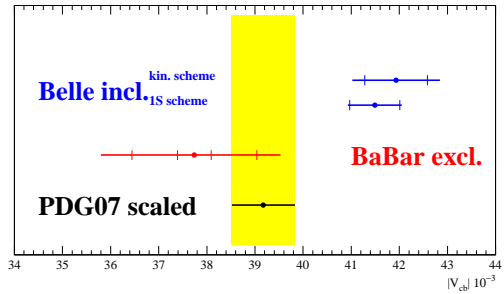}
\caption{\label{fig:summary-vcb}
Comparison of the discussed results on \VCB\ and the current world average given by the
HFAG as preliminary result for PDG07 \protect\cite{HFAG}.
The world average is scaled with the value $\mathcal{F}(1)=0.91$ \protect\cite{pdg2006} to be
comparable with the inclusive results.}
\end{figure}

For \VCB\ the precision has improved and the uncertainties from theoretical
calculations start to dominate the results together with the uncertainty of $\Gamma_{SL}$.
However, with the given precision it becomes possible to use theory calculations not
only as input (like normalizations of form factors in exclusive analyses) but also test the
predictions for consistency (as for global fits to kinematic moments).
So far fits to calculations with different ansatzes agree very well.

Overall the knowledge on \VCB\ has reached a level of about 2\%. The most recent results
(Fig. \ref{fig:summary-vcb}) are
\begin{eqnarray}
|V_{cb}| &=& (41.49\pm0.52_{fit}\pm0.20_{\tau_B}) \cdot 10^{-3} \nonumber \\
|V_{cb}| &=& (37.74 \pm 0.35_{stat} \pm 1.25_{syst} \,^{+1.23}_{-1.44} \,_{\Gamma_{SL}} ) \cdot 10^{-3} \nonumber
\end{eqnarray}
from fits to moments of inclusive analyses and exclusive decays $B^0 \to D^{*-}\ell^+ \nu$ respectively.

\section*{Acknowledgments}

I would like to thank all the colleagues from the Belle and \babar\ Collaborations,
especially Christoph Schwanda and David Lopes Pegna, for their help and support during the
preparation of this talk.

I would like also to thank all the organizers, in particular Peter Krizan and his local team.
They managed to have everything arranged in such a perfect manner that this conference was an
untroubled and inspiring discussion of a rich scientific program within the stimulating scenery
of lake Bled.

Finally I would like to thank the BMBF, Germany, for the funding of this work.


\begin{thebibliography}{99} % Use for 10-99 references

\bibitem{DSexcl:caprini-ff}
I.\ Caprini, L.\ Lellouch, M.\ Neubert,
Nucl.\ Phys.\ B {\bf 530}, 153 (1998). 

\bibitem{DSexcl:hashimoto-LQCD}
S.\ Hashimoto {\em et al.},
Phys.\ Rev.\ D {\bf 66}, 014503 (2002). 

\bibitem{DSexcl:babar-e}
The \babar\ Collaboration,
B.\ Aubert {\em et al.},
Phys.\ Rev.\ D {\bf 74}, 092004 (2006). 

\bibitem{DSexcl:babar-comb}
The \babar\ Collaboration,
B.\ Aubert {\em et al.},
hep-ex/0607076 (2006).

\bibitem{DSS:D0.ratio}
The D0 Collaboration,
V.\ M.\ Abazov {\it et al.},
Phys.\ Rev.\ Lett.\ {\bf 95}, 171803 (2005). 

\bibitem{DSS:belle.D1-nonres}
The Belle Collaboration,
K.\ Abe {\em et al.},
Phys.\ Rev.\ Lett.\ {\bf 94}, 221805 (2005). 

\bibitem{CombBR:babar}
The \babar\ Collaboration,
B.\ Aubert {\em et al.},
hep-ex/0703027

\bibitem{pdg2006}
Particle Data Group, 
W.-M. Yao {\em et al.},
Journal of Physics G {\bf 33}, 1 (2006).

\bibitem{HQE:kin-scheme}
P.\ Gambino and N.\ Uraltsev,
Eur.\ Phys.\ J.\ C {\bf34}, 181 (2004). 

\bibitem{HQE:1s-scheme}
C.\ W.\ Bauer, Z.\ Ligeti, M.\ Luke, A.\ V.\ Manohar and M.\ Trott
Phys.\ Rev.\ D {\bf 70}, 094017 (2004). 

\bibitem{HQE:belle-ee}
The Belle Collaboration,
P.\ Urquijo {\em et al.},
Phys.\ Rev.\ D {\bf 75}, 032001 (2007). 

\bibitem{HQE:belle-mhad}
The Belle Collaboration,
C.\ Schwanda {\em et al.},
Phys.\ Rev.\ D {\bf 75}, 032005 (2007). 

\bibitem{HQE:belle-egamma}
The Belle Collaboration,
K.\ Abe {\em et al.},
hep-ex/0508005 (2005).

\bibitem{HQE:belle-params}
The Belle Collaboration,
K.\ Abe {\em et al.},
hep-ex/0611047 (2005).

\bibitem{HQE:other-ee}
The \babar\ Collaboration,
B.\ Aubert {\em et al.}
Phys.\ Rev.\ D {\bf 69}, 111104 (2004). 

The CLEO Collaboration,
A.\ H.\  Mahmood {\em et al.}
Phys.\ Rev.\ D  {\bf 70}, 032003 (2004). 

The DELPHI Collaboration,
J.\ Abdallah {\em et al.},
Eur.\ Phys.\ J.\ C {\bf 45}, 35 (2006). 

\bibitem{HQE:other-mhad}
The \babar\ Collaboration,
B.\ Aubert {\em et al.}
Phys.\ Rev.\ D {\bf 69}, 111103 (2004). 

The CLEO Collaboration,
S.\ E.\ Csorna {\em et al.},
Phys.\ Rev.\ D {\bf 70}, 032002 (2004). 

The DELPHI Collaboration,
J.\ Abdallah {\em et al.},
Eur.\ Phys.\ J.\ C {\bf 45}, 35 (2006). 

The CDF Collaboration,
D.\ Acosta {\em et al.},
Phys.\ Rev.\ D {\bf 71}, 051103 (2005). 

\bibitem{HQE:buchmueller-params}
O.\ Buchm\"uller, H.\ Fl\"acher,
Phys.\ Rev.\ D {\bf 73}, 073008 (2006). 

\bibitem{HFAG}
The Heavy Flavour Averaging Group,
E.\ Barberio {\em et al.}, \\
\verb|http://www.slac.stanford.edu/xorg/hfag/|

\end{thebibliography}
\end{document}
%
% ****** End of file template.aps ******